# A Synchrotron in the TEM: Spatially Resolved Fine Structure Spectra at High Energies

James L Hart[1,2], Andrew C Lang[1,3], Yuanyuan Li[4], Kanit Hantanasirisakul[1,5], Anatoly I Frenkel[4,6], Mitra L Taheri[1,2]

1. Department of Materials Science and Engineering, Drexel University, Philadelphia, PA, USA
2. Department of Materials Science and Engineering, Johns Hopkins University, Baltimore, MD, USA
3. American Society for Engineering Education Postdoctoral Fellowship, Washington, DC, USA
4. Department of Materials Science and Chemical Engineering, Stony Brook University, Stony Brook, NY, USA
5. A.J. Drexel Nanomaterials Institute, Drexel University, Philadelphia, PA, USA
6. Division of Chemistry, Brookhaven National Laboratory, Upton, NY, USA

**Fine structure analysis of core electron excitation spectra is a cornerstone characterization technique across the physical sciences. Spectra are most commonly measured with synchrotron radiation and X-ray spot sizes on the µm to mm scale. Alternatively, electron energy loss spectroscopy (EELS) in the (scanning) transmission electron microscope ((S)TEM) offers over a 1000 fold increase in spatial resolution, a transformative advantage for studies of nanostructured materials. However, EELS applicability is generally limited to excitations below ~2 keV, i.e., mostly to elements in just the first three rows of the periodic table. Here, using state-of-the-art EELS instrumentation, we present nm resolved fine structure EELS measurements out to an unprecedented 12 keV with signal-to-noise ratio rivaling that of a synchrotron. We showcase the advantages of this technique in exemplary experiments.**

The fine structure of a core level excitation spectrum was first measured *via* X-ray absorption spectroscopy (XAS) in the early 1900s, with the 1924 Nobel prize awarded to Siegbahn for his pioneering work. Since then, fine structure measurements have emerged as an essential characterization method for physics, chemistry, biology, and materials science [1]. Observed near the onset of the absorption edge, the fine structure is related to the local density of states and the atomic site symmetry, allowing detailed chemical bonding analysis [2]. From ~50 to 1000 eV past the edge, the extended fine structure is related to the local environment of the excited atom, allowing the determination of element-specific bond lengths, coordination numbers, coordinating species, and their dynamics [3,4]. With XAS, the X-ray absorption near-edge structure (XANES) and extended X-ray absorption fine structure (EXAFS) can be probed with excellent signal-to-noise ratio (SNR) out to several tens of keV, enabling analysis of the full periodic table. As an alternative to XAS, in the 1940s Hillier used inelastic electron scattering to probe core level excitations in the (S)TEM [5]. Analogous to XANES and EXAFS, EELS can probe the energy loss near edge structure (ELNES) and the extended energy loss fine structure (EXELFS) [6]. While X-ray spot sizes are mostly on the µm scale, the electron probe of a STEM can be focused to sub Å dimensions, offering more than a 1000 fold increase in spatial resolution (Fig. 1A) [7].

This permits, for example, local mapping of chemical bonding around individual atomic defects [8]. EELS offers the additional advantage of correlating measurements from the meV scale out to the keV scale with a single instrument, encompassing phonons, bandgaps, plasmons, and core-loss edges of both light and heavy elements [6,9]. Conversely, individual synchrotron beamlines are restricted to specific energy windows, *e.g.* tender X-ray beamlines can only probe ~1 – 5 keV (Fig. 1A). Despite these benefits, a challenge with EELS is that the scattering cross section rapidly decreases with energy, leading to SNR limitations for excitations above ~2 keV. This is particularly problematic for EXELFS, which requires wavenumber ($k$) range from 0 to at least 8 – 10 Å$^{-1}$ (or about 300 – 400 eV post-edge energy range). Practically speaking, this $k$ range necessitates high SNR measurements. Consequently, EXELFS (and to a lesser extent ELNES) analysis has been limited to low energies [10–13], restricting the study of many important edges, *e.g.* the $K$-edges of 3$d$ transition metals or the $L$-edges of period 5 and 6 elements.

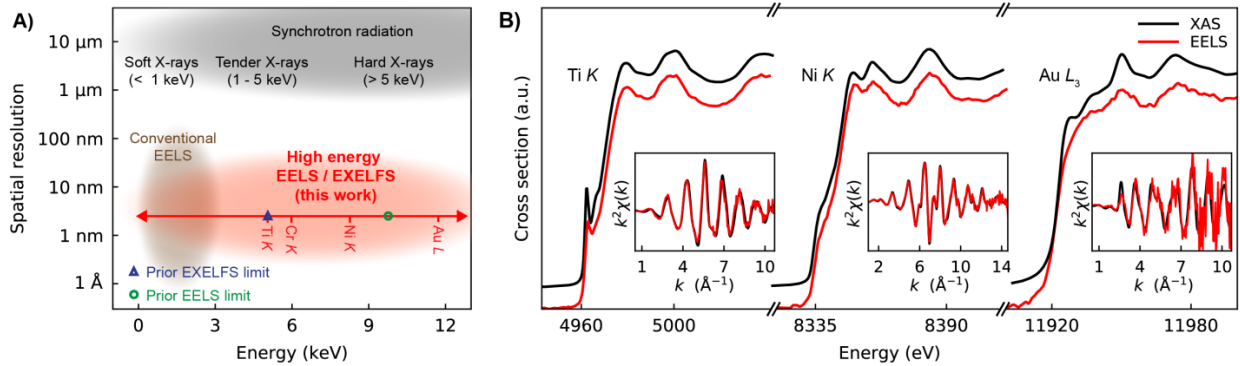

**Figure 1. (A)** Plot of spatial resolution and energy range for conventional EELS, XAS, and the work presented herein. The prior EXELFS limit comes from ref. [14] and the prior EELS limit comes from ref. [13]. **(B)** EELS and XAS comparison for the $K$-edge of Ti, the $K$-edge of Ni, and $L_3$-edge of Au. The insets show EXELFS versus EXAFS comparisons.

In this report, we use advanced instrumentation to perform localized EELS fine structure measurements with high SNR out to a record 12 keV (Fig. 1B). Past high energy EELS experiments used indirect detection sensors which significantly reduce the measured SNR [13,14]. We take advantage of modern direct detection electron sensors which enable counting of individual incident electrons [15]. This method provides nearly shot noise limited acquisition and has previously been used to increase the SNR in time-resolved and low dose EELS experiments [16–19]. Additionally, recent work has highlighted the importance of optimizing the coupling between the TEM and spectrometer for high energy measurements [13,20,21], and here, we use a customized lens configuration to improve signal collection and reduce high energy artifacts (Supplementary Information). With this approach, we obtain high energy EELS data with SNR comparable to synchrotron XAS measurements. We showcase the advantages of this new ability in two key examples. First, taking advantage of the nanoscale spatial resolution of EELS, we perform localized EXELFS measurements on a nanolaminate of crystalline Ni and the bulk metallic glass (BMG) $Zr_{65}Cu_{17.5}Ni_{10}Al_{7.5}$. Through EXELFS analysis, we reveal preferential nearest neighbor bonding within the BMG which could not have been detected with XAS or conventional STEM EELS techniques. Second, taking advantage of the increased spectral range of EELS, we provide a comprehensive understanding of surface chemistry in a newly-discovered 2D material (belonging to the MXene family) by correlating EELS data on core-loss edges from < 1 keV out to 6 keV, spanning the soft to hard X-ray regimes in a single instrument.

**Results**

First, we demonstrate the ability to measure high energy edges with synchrotron quality SNR using a focused electron probe. Figure 1B presents side-by-side comparisons of EELS and XAS for several reference edges: the *K*-edge of Ti (5 keV), the *K*-edge of Ni (8.3 keV), and the $L_3$-edge of Au (11.9 keV). Good agreement is observed between the two techniques, though there is a slight reduction in energy resolution, which we describe further in the Discussion section. To compare the extended edge data, we follow the standard procedure of isolating and normalizing the extended edge signal and then converting the excited (photo)electron energy to its wave number, obtaining $\chi(k)$ [1]. The insets compare the $k^2$ weighted $\chi(k)$ data for each edge. Excellent agreement between EXELFS and EXAFS is evident. Our Au $L_3$-edge measurement at 11.9 keV is, to our knowledge, the highest energy EELS (and EXELFS) result reported to date. The prior EELS limit was the W $L_3$-edge at 10.2 keV measured by MacLaren *et al* [13], though the SNR was too low for EXELFS analysis. The prior EXELFS high energy mark was the Ti *K*-edge at 5 keV measured by Blanche and Hug [14]. By extending the accessible energy range of EXELFS out to >10 keV, we can now access important excitations such as all the 3*d* transition metal *K*-edges and other high *Z* elements' *L* and *M* edges. We also note that, to achieve adequate SNR, Blanche and Hug used an extraordinarily large beam current of 400 nA, which, owing to sample damage considerations, precludes the measurement of all materials other than extremely stable metals and ceramics [22]. In contrast, for our measurement of the T *K*-edge we use a 2 nA beam (Supplementary Information), enabling the analysis of a much wider array of materials.

Having demonstrated high quality EELS at record high energies, we next consider the advantages offered by this technique in several archetypal applications. With respect to EXAFS measurements – which are inherently ensemble-averaging over μm length scales [23] – EXELFS provides a dramatic improvement in spatial resolution. In comparison to other localized (S)TEM techniques such as imaging, diffraction, tomography, or conventional EELS elemental mapping, EXELFS is uniquely able to quantify short range order (SRO) – defined here as local bond lengths, coordination numbers, and identities of coordination species – in disordered, multi-component materials. This new ability (local analysis of SRO) is critical for the study of complex materials possessing disorder and nanoscale heterogeneity which have, to date, eluded accurate characterization [23,24]. As a case study, we investigate local structure in a laminate of crystalline Ni and the BMG $Zr_{65}Cu_{17.5}Ni_{10}Al_{7.5}$ (Fig. 2) [25]. Such laminates offer high strength with improved ductility compared to single-phase BMGs [26]. However, crystallization within the BMG during processing may hinder mechanical properties, and as such, it is important to quantify SRO within the BMG after the laminate is formed [25]. To this end, we study variation in the local environment of Ni through EXELFS measurements of $\chi(R)$, which is the Fourier transform of $\chi(k)$ and is closely related to the radial distribution function of the excited atom. With EXAFS, Ni *K*-edge measurements would yield ensemble averaged data, mixing together the signals from the crystalline Ni and BMG structures [27]. Conversely, with STEM-EXELFS, we are able to selectively probe both the crystalline and BMG phases (Fig. 2A-C). In contrast to the crystalline Ni, the BMG Ni *K*-edge $|\chi(R)|$ shows little intensity beyond ~3 Å, indicative of its amorphous structure (Fig. 2C). Comparison of the measured $|\chi(R)|$ from BMG regions 1-4 (Fig.

2B) reveals no significant differences in BMG structure at interfaces versus the phase interior (Supplementary Information). Interestingly, while the BMG is primarily composed of Zr, EXELFS fitting shows that Ni is mostly bonded with Ni and/or Cu, indicating a significant degree of bond ordering (Fig. 2D, Supplementary Information). Such preferential bonding can strongly influence the BMG's mechanical properties [28], and we stress that this result could not have been directly obtained with EXAFS (owing to the limited spatial resolution) or conventional STEM techniques, which cannot directly probe SRO in multi-component, disordered samples. These nanoscale STEM-EXELFS measurements have implications for a number of emerging material systems which possess disorder, *e.g.* high entropy alloys (which are *chemically* disordered) and amorphous magnets.

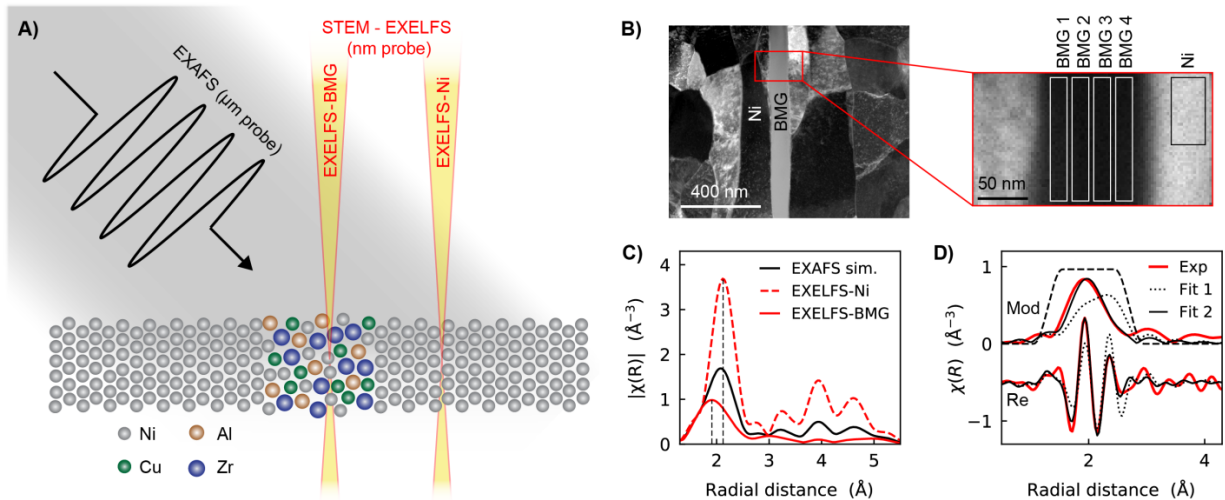

**Figure 2. (A)** Schematic showing the structure of the laminate and the improved spatial resolution of STEM-EXELFS compared to EXAFS. **(B)** STEM annular dark field (ADF) image showing the laminate. The magnified region shows a Ni *K*-edge intensity map, taken from a STEM-EXELFS spectrum image. **(C)** Comparison of the Ni *K*-edge EXELFS $|\chi(R)|$ ($k^2$ weighted) for crystalline Ni and the BMG, with the BMG data being the average of regions 1-4 shown in (B). The EXAFS simulation is the average of the EXELFS Ni and BMG measurements, representing an ensemble averaged measurement acquired with a µm scale X-ray spot size. **(D)** Fits to the BMG EXELFS data shown in (C). The dashed black line shows the fit window in *R*-space. In Fit 1, the relative amount of Ni-Zr and Ni-Ni nearest neighbor (n.n.) bonding is fixed and proportional to the nominal BMG composition. In Fit 2, we allow the degree of Ni-Zr versus Ni-Ni n.n. bonding to vary, and we find a strong preference for Ni-Ni bonding, see Supplementary Information.

Next, we consider the improved spectral range of EELS; while synchrotron beamlines used for XAS can only access specific energy windows (for instance, hard X-ray beamlines used for EXAFS can generally only access >5 keV), a single EELS spectrometer can access from < 1 eV out to >10 keV. Hence, EELS allows measurement of both high Z and low Z elements, which is not possible with a single synchrotron experiment. Additionally, analysis of a single element can often benefit from the measurement of both high energy and low energy edges, which may selectively probe different unoccupied orbitals. To demonstrate these advantages, we study surface chemistry evolution in the 2D $Cr_2TiC_2T_x$ MXene (T = surface termination), where we measure from ~1 eV out to 6 keV in a single experiment (Fig. 3A,B). The MXene family shows state-of-the-art performance in fields such as energy storage and electromagnetic interference shielding [29,30]. Owing to current synthesis methods, their surfaces are terminated with a mix

of –OH, –F, and –O species. Terminations can strongly alter the MXene properties [18], and as such, it is critical to quantify and control the surface chemistry. We perform *in situ* annealing of $Cr_2TiC_2T_x$ to thermally induce surface termination loss, which we study *via* Cr *K*-edge EELS (the surfaces of this MXene are entirely composed of Cr) (Fig. 3A-C). In the Cr *K*-edge $|\chi(R)|$ data, the first peak at $R \sim 1.4$ Å is related to Cr-C and Cr-T bonding, and the decrease in this peak with annealing indicates termination loss (Fig. 3D). EXELFS fitting indicates a 45±8% decrease in the termination concentration after 600 °C annealing, which has significant implications for the electronic structure of this material (Fig. 3E and Supplementary Information) [31,32]. However, owing to the similar weight of F and O, the EXELFS modeling cannot differentiate between –OH, –F or –O loss. Looking to lower energies, EELS analysis of the O *K*- and F *K*-edges reveals that the measured decrease in termination concentration is mainly due to loss of –F (Fig. 3F). We highlight that a more complete understanding of the MXene surface chemistry evolution was only gained by correlating the Cr *K*-edge at 6 keV with multiple edges below 1 keV which are inaccessible to (hard) EXAFS beamlines. Beyond this example, the improved (low) energy range of EELS holds wider potential for the study of nitrides, oxides, fluorides, or any other material which contains both light and heavy elements. Additionally, correlation of high energy EXELFS with low loss EELS is an exciting avenue. For example, in plasmonic nanoparticles, local structure could be measured with EXELFS and then correlated with the material permittivity as measured with low loss EELS [6].

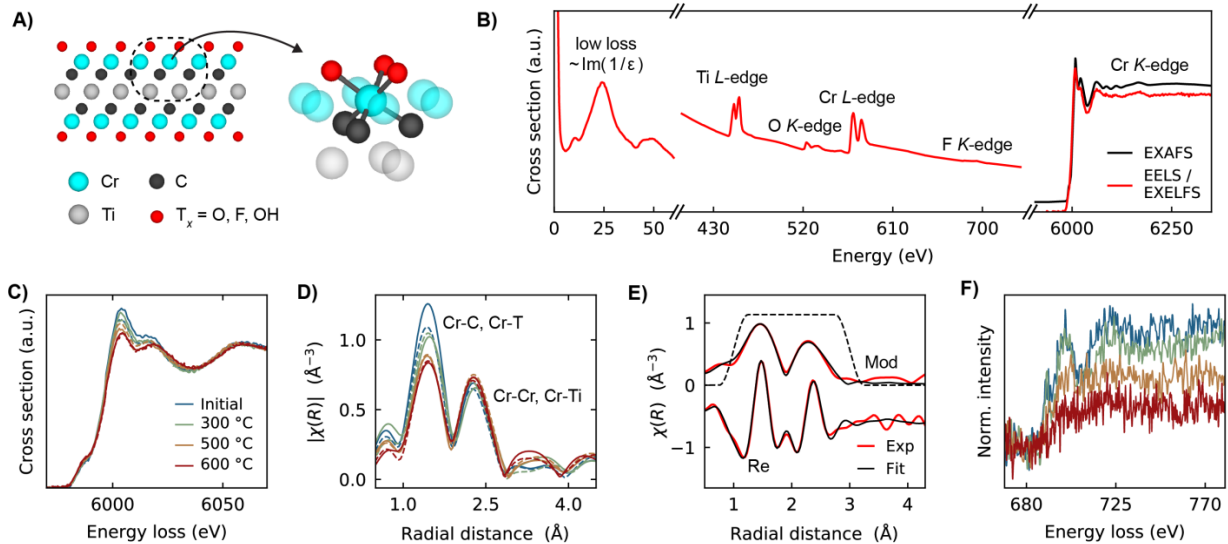

**Figure 3. (A)** Schematic showing the $Cr_2TiC_2T_x$ structure in cross-section (left), and a magnified view of the local Cr environment (right). T represents the mixed surface termination. **(B)** Comparison of the EELS / EXELFS energy range with the EXAFS energy range. ε is the complex permittivity of the material. **(C)** Evolution of the Cr *K*-edge ELNES. Data was aligned based on the edge onset. Two areas of the sample were measured, differentiated by the solid and dashed lines. All measurements were made at room temperature after annealing. **(D)** Evolution in the Cr *K*-edge $|\chi(R)|$ ($k^2$ weighted) with annealing. **(E)** Example fit to a dataset from (C). The dashed black line shows the fit window in *R*-space. **(F)** Evolution of the F *K*-edge intensity with annealing, normalized based on the Cr *L*-edge intensity.

## Discussion

While we have demonstrated high SNR EELS with nanoscale spatial resolution up to 12 keV, it is worth considering the high energy limitations still facing this technique. Principally, SNR continues to be a major challenge; the spatial resolution of our data is not limited by the electron probe size, but rather by sample drift given the long exposures needed to gain sufficient SNR. This challenge could be partially addressed through further improvement in detector technology and TEM-spectrometer coupling, as well as more stable TEM stages. Energy resolution presents another challenge. The microscope used for these experiments has an energy resolution of ~1.5 eV at 0 eV energy loss, but at higher energy losses, the energy resolution is considerably lower. This is evident in the broadening of the Ti $K$-edge and the Au $L_3$-edge fine structure (Fig. 1B). This decrease in resolution at high energy has been observed previously [13], and may be related to the large (lateral) momentum transfer typically associated with high energy loss scattering [6]. Additionally, the calibration of the energy loss axis is challenging at high energies, which in this study prevented quantitative chemical shift analysis of the MXene Cr $K$-edge ELNES with annealing (Fig. 3C). Both of these issues (energy resolution and calibration) must be addressed for quantitative ELNES analysis at high energy in the (S)TEM.

In conclusion, we utilized direct detection EELS to perform nanoscale EELS and EXELFS measurements with record SNR out to a record 12 keV. Our approach enables highly localized measurements of SRO, with three orders of magnitude smaller spot sizes compared to those previously possible with EXAFS. This ability is relevant for the characterization of disordered, nanostructured and/or heterogeneous materials, which include many important material systems ranging from catalytic nanoparticles to 2D materials to metal organic frameworks. Moreover, in our experiments we demonstrated that EELS, as a single instrument, is capable of providing spectral analysis from < 1 eV out to >10 keV. Collection of spectral data from such disparate energy scales on a single instrument has wide implications for locally correlating structure, chemistry, elemental composition, and functional properties.

## Acknowledgements:

The authors thank Prof. Ian MacLaren and Rebecca Cummings for helpful discussions, particularly with respect to electron optics at high energy and EXELFS fitting. The authors thank Prof. Suveen Mathaudhu and Sina Shahrezaei for providing the BMG sample, Prof. Yury Gogotsi for supporting the MXene work, and Prof. Babak Anasori for providing $Cr_2TiAlC_2$ MAX precursor for MXene synthesis. JL Hart, AC Lang, and ML Taheri acknowledge funding from the National Science Foundation (NSF) MRI award #DMR-1429661, and thank the Drexel Centralized Research Facilities for supporting electron spectroscopy measurements. Y Li and AI Frenkel gratefully acknowledge support for this work by the U.S. Department of Energy (DOE), Office of Basic Energy Sciences under Grant No. DE-FG02-03ER15476. This research used beam line 7-BM (QAS) of the National Synchrotron Light Source II, a U.S. DOE Office of Science User Facility operated for the DOE Office of Science by Brookhaven National Laboratory under contract no. DE-SC0012704. Beam line operations were supported in part by the Synchrotron Catalysis Consortium (U.S. DOE, Office of Basic Energy Sciences, grant no. DE-SC0012335). K Hantanasirisakul is supported by the U.S. DOE, Office of Science, Office of Basic Energy Sciences, grant no. DE-SC0018618.

**Supplementary Information:**

*Sample preparation:*

The BMG sample was prepared as in ref. [25], and a $Ga^+$ FIB was used to prepare the TEM specimen. The MXene sample was prepared as in ref. [34], and MXene solution was drop-cast onto a lacey carbon grid.

*EXELFS acquisition:*

All electron microscopy and spectroscopy were performed on a JEOL 2100F instrument with an accelerating voltage of 200 kV and a Schottky emitter. EELS was collected with a Gatan Imaging Filter (GIF) Quantum equipped with a Gatan K2 IS camera operated in electron counting mode [16]. The GIF high SNR aperture (5 mm) was used for all experiments. The dispersion was nominally set to 0.5 eV/channel, but, after EXELFS calibration on the Ti, Ni and Au reference samples, the dispersion for all edges was reset to 0.485 eV/channel. A custom post-specimen lens configuration was used, with a collection semi-angle that varied from 73 mR at 5 keV energy loss to 56 mR at 10 keV energy loss (Supplementary Figure 1). A power-law background subtraction was performed. For the MXene experiment, the microscope was operated in TEM mode to spread the electron dose, and for all other measurements, the microscope was operated in STEM mode with a convergence semi-angle of 16 mR. For all EXELFS measurements, multiple acquisitions of ~60 s were performed, and individual spectra were then aligned and summed.

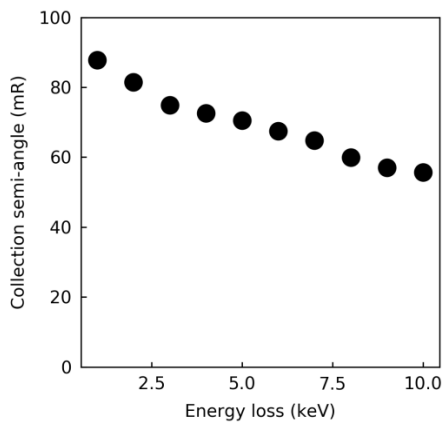

**Supplementary Figure 1.** Evolution of the collection semi-angle with energy loss. The collection angle was measured using a reference CBED pattern while reducing the microscope high tension in STEM mode. As the high tension was reduced, the C1 condenser lens was adjusted to keep the STEM probe focused on the specimen. EELS measurements performed at greater than 10 keV energy loss were accomplished by increasing the TEM high tension.

For the Ti $K$-edge measurement shown in Fig. 1, the beam current was 2 nA and the total acquisition time was 600 s, giving a total dose of $7.6 \times 10^{12}$ electrons. The sample had a thickness of $t/\lambda = 1.1$, where $t$ is the sample thickness and $\lambda$ is the inelastic mean free path. For the Ni $K$-edge measurement shown in Fig. 1, the beam current was 6 nA and the total acquisition time was 1200 s, giving a total dose of $4.5 \times 10^{13}$ electrons. The sample had a thickness of $t/\lambda =$

1.0. For the Au $L_3$-edge measurement shown in Fig. 1, the beam current was 8 nA and the total acquisition time was 3000 s, giving a total dose of $1.5 \times 10^{14}$ electrons. The sample had a thickness of $t/\lambda = 1.2$. For all of these experiments, data was collected as the STEM probe continuously rastered across an area of ~100×100 nm$^2$

For the spatially resolved mapping of the Ni / BMG laminate, the STEM-EXELFS spectrum image (SI) had a step size of 2 nm and total dwell time per pixel of 0.3 s, achieved with a multi-frame SI acquisition with 30 passes and 0.01 s per pass. Seven multi-frame SIs were acquired and summed. The STEM probe current was 8 nA, and the average sample thickness across the SI area was $t/\lambda = 0.55$.

For the MXene Cr $K$-edge measurements, the TEM beam current was 7 nA, and each EXELFS measurement had a total acquisition time of 600 s, giving a total dose of $2.6 \times 10^{13}$ electrons for each measurement. For the 8 areas sampled, the thicknesses ranged from $t/\lambda = 0.7$ up to 1.1. Annealing was performed with the Gatan 626 hot stage.

*EXAFS acquisition:*

The Cr $K$-edge X-ray absorption spectra were collected at 7-BM (QAS) beamline, National Synchrotron Light Source II, Brookhaven National Laboratory (BNL). Samples were prepared by vacuum filtration of MXene solution. The data were collected in transmission mode. For Ni $K$-edge, Ti $K$-edge, and Au $L_3$-edge spectra of foils, all the data were taken from the web depository of synchrotron catalysis consortium at BNL at http://you.synchrotron.edu/scc2.

*EXELFS and EXAFS data processing:*

Prior to summation, individual EELS datasets were processed to remove 'hot pixels' (pixels with aberrant values due to issues with the counting software). After summation, plural scattering was removed with a Fourier ratio deconvolution as implemented in Gatan Digital Micrograph v.3. EELS and XAS data were normalized using the Athena software package [33]. Shortly, low order polynomials were fit to the pre- and post-edge regions, allowing extraction of the normalized cross-section. The absolute energy calibration for EELS at high energy is poor, and all EELS edges were aligned to the XAS data. All EXELFS fitting was performed with the Artemis software package [33].

*Additional EXELFS data and fitting:*

A comparison of the $|\chi(R)|$ from BMG regions 1-4 is shown in Supplementary Figure 2.

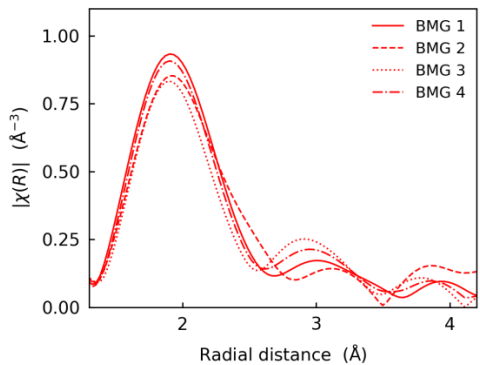

**Supplementary Figure 2.** EXELFS comparison of BMG regions 1-4 as shown in the manuscript Fig. 2. The $|\chi(R)|$ data is $k^2$ weighted.

For the BMG fitting, we focused on the nearest neighbor bonding and only considered single scattering paths. We began with a model where the excited Ni atom is bonded to neighboring Ni, Cu, Zr, and Al atoms, with the starting bond distances determined by the sum of the metallic radii of Ni and the neighboring species. With this model, we sought to fit $S_0^2$ (the amplitude reduction factor) and a bond distance, $R$, for each scattering path. With this approach, the fitted value of $S_0^2$ acts as a proxy for the likelihood of Ni bonding to the given neighboring species. Owing to the similar $Z$ of Ni and Cu, the EXELFS signals arising from the Ni-Ni and Ni-Cu scattering paths are similar. In an effort to simplify the fit, we thus removed Cu from the model. The fitted $S_0^2$ for the Ni-Ni path is then assumed to correlate with the likelihood Ni-Ni and/or Ni-Cu bonding. Initial fits consistently showed the $S_0^2$ for the Ni-Al path to be ~0, which is not surprising given the low concentration of Al in the alloy. The Al-Ni path was therefore removed from the fit. Our final fit then included 4 free parameters: $S_0^2$ for Ni-Ni, $S_0^2$ for Ni-Zr, $R$ for Ni-Ni, and $R$ for Ni-Zr. We assumed that the in the amorphous structure $\sigma^2$ (the bond length mean square deviation) was ~0.01 Å$^2$, and we did not fit an energy offset. The resulting best fit is shown in the manuscript Fig. 2D, and the fit parameters are given in Supplementary Table 1.

**Supplementary Table 1.** EXELFS fit details to the Ni $K$-edge within the BMG phase. In this fit, $S_0^2$ correlates with the Ni-Ni (or Ni-Zr) coordination number, and hence the likelihood of Ni-Ni (or Ni-Zr) bonding within the amorphous structure. $R$ is the bond length.

| Fit parameters | |
| --- | --- |
| $k$ range | 3 – 10 |
| $R$ range | 1.3 – 2.7 |
| Independent points | 6.11 |
| Number of fitted parameters | 4 |
| **Fit statistics** | |
| $\chi^2$ | 35.48 |
| Reduced $\chi^2$ | 16.83 |
| $R$-factor | 0.09953 |
| **Fitted values** | |
| $S_0^2$ Ni-Ni | 2.971 ± 0.6952 |
| $S_0^2$ Ni-Zr | 1.441 ± 1.039 |
| $R$ Ni-Ni | 2.466 ± 0.025 Å |
| $R$ Ni-Zr | 2.584 ± 0.056 Å |

We note that the large values of $S_0^2$ obtained here are unphysical, which is the result of setting the bond multiplicity of Ni-Ni and Ni-Zr to 1. The main result from this analysis is that the fitted $S_0^2$ for Ni-Ni is greater than that of Ni-Zr, indicating that Ni is more likely to be bonded with Ni and/or Cu than with Zr, despite the nominal atomic concentration of Zr being higher than the combined concentration of Ni and Cu. We note the large errors in the fitted values $S_0^2$; however, regardless of the set value of $\sigma^2$ or the set value of $\Delta E_0$ (which were not fit, but were manually varied) the fitted $S_0^2$ for Ni-Ni was consistently greater than that of Ni-Zr. We also manually adjusted the starting Ni-Ni and Ni-Zr $R$ values, and while the final fitted structure was sensitive to the initial conditions, the Ni-Ni $S_0^2$ was consistently greater than that of Ni-Zr. We additionally performed a fit with just Ni-Ni scattering, and then another with only Ni-Zr scattering. The fit with only Ni-Ni scattering provided much better agreement with the experimental data, further supporting our conclusion of preferential Ni-Ni and Ni-Cu bonding.

For the $Cr_2TiC_2T_x$ MXene EXELFS fitting, we treated all termination species as O atoms (given the similar $Z$ of O and F), and we assumed that the surfaces were initially fully terminated. With annealing, we assume the only change in structure is the loss of surface terminations (meaning a decrease in the Cr coordination), which we capture by allowing $S_0^2$ for the Cr-T scattering path to vary between each dataset. This was achieved by defining an $S_0^2$ reduction factor (RF) for each Cr $K$-edge measurement. All of the other fitted parameters were kept constant across all datasets: $S_0^2$, the initial amplitude reduction factor, $R_T$, the Cr-T bond distance, $\Delta R$, the percent change in the Cr-C, Cr-Cr, and Cr-Ti bond lengths from their nominal values, $\sigma_T^2$, the bond length mean square deviation for Cr-T bonding, and $\sigma_L^2$, the 'lattice' bond length mean square deviation assigned to Cr-C, Cr-Cr, and Cr-Ti bonds. Additionally, energy offsets ($\Delta E_0$) were fit to each dataset. Fit details are given in Supplementary Table 2.

**Supplementary Table 2.** EXELFS fit details to the Cr $K$-edge of $Cr_2TiC_2T_x$ (MXene). All of the different parameters in the table are defined within the Supplementary text.

| Fit parameters | |
|---|---|
| $k$ range | 2.5 – 9 |
| $R$ range | 1 – 3 |
| Independent points | 64.5 |
| Number of fitted parameters | 20 |
| Number of datasets | 8 |
| **Fit statistics** | |
| $\chi^2$ | 979.5 |
| Reduced $\chi^2$ | 22.01 |
| $R$-factor | 0.01968 |
| **Fitted values** | |
| $S_0^2$ | 0.7921 ± 0.03836 |
| RF initial 1 | 1 (set value, not fit) |
| RF initial 2 | 0.794 ± 0.087 |
| RF 300 °C 1 | 0.782 ± 0.107 |
| RF 300 °C 2 | 0.761 ± 0.077 |
| RF 500 °C 1 | 0.583 ± 0.065 |
| RF 500 °C 2 | 0.596 ± 0.060 |
| RF 600 °C 1 | 0.516 ± 0.055 |
| RF 600 °C 2 | 0.487 ± 0.093 |
| $R_T$ | 1.986 ± 0.004 Å |
| $\Delta R$ | -0.00821 ± 0.0019 |
| $\sigma^2_T$ | 0.00334 ± 0.00084 Å$^2$ |
| $\sigma^2_L$ | 0.01695 ± 0.00094 Å$^2$ |